# Superconductivity of topological insulator $Bi_2Se_3$ at high pressures

P. P. Kong, J. L. Zhang, S. J. Zhang, J. Zhu, Q. Q. Liu, R. C. Yu, Z. Fang, C. Q. Jin[*]

Beijing National Laboratory for Condensed Matter Physics, and Institute of Physics, Chinese Academy of Sciences, Beijing, 100190, China

W. G. Yang

High Pressure Synergetic Consortium (HPSynC) and High Pressure Collaborative Access Team (HPCAT), Geophysical Laboratory, Carnegie Institution of Washington, Argonne, Illinois 60439, United States

X. H. Yu, J. L. Zhu, Y. S. Zhao

Los Alamo s Neutron Science Center (LANSCE), Los Alamos National Laboratory, Los Alamos, New Mexico 87545, United States

**Abstract**

The pressure-induced superconductivity and structural evolution for Bi2Se3 single crystal have been studied. The emergence of superconductivity with onset transition temperature (Tc) about 4.4K is observed around 12GPa. Tc increases rapidly to the highest 8.5K at 16GPa, decreases to 6.5K at 21GPa, then keep almost constant. It is found that Tc versus pressure is closely related to the carrier density which increases by more than two orders of magnitude from 2GPa to 23GPa. High pressure synchrotron radiation measurements reveal structure transitions occur around 12GPa, 20GPa, and above 29GPa, respectively. A phase diagram of superconductivity versus pressure is obtained.

Topological insulator (TI) with a bulk band gap but gapless edge state protected by time-reversal symmetry is a current frontier in condensed matter physics, attracting worldwide research interests.[1~8] There are many useful properties expected for topological compounds, ranging from new physics of Majorana Fermion,[9] prospective applications of topological computer[10], to exotic topological superconductors.[11] Recently a group three dimensional TIs of $Bi_2Te_3$, $Bi_2Se_3$, etc., are predicated, consequently observed by experiments such as angle-resolved photoemission spectroscopy (ARPES).[4, 7, 11-13]

Similar to TI, topological superconductor has a full pairing gap in the bulk and gapless edge state consisting of elusive Majorana fermions.[14] When topological insulator combines with magnetic or superconducting materials, Majorana state can be realized at its interface. Several application[12, 15,16] of interface between a topological insulator and a superconductor also have been proposed. The superconductivity of $Bi_2Se_3$ was observed via copper intercalating into the van der Waals gaps between the quintuple layers,[17] while $Bi_2Te_3$ becomes superconductive at pressure.[18,] Pressure is much effective in generating or tuning superconductivity by modifying electronic structure through directly changing interatomic distance without introducing defects or impurities comparing with chemical doping. Here we report studies on the effects of pressure on $Bi_2Se_3$, We found that $Bi_2Se_3$ becomes superconducting at its high pressure phase.

Single crystal $Bi_2Se_3$ was grown by Bridgeman method. The details are described in Ref. 18. High purity elements Bi (99.999%) and Se (99.999%) were mixed in a molar ratio of 2:3, ground and pressed into pellets, then loaded into a quartz Bridgeman ampoule, which was evacuated and sealed. The ampoule was placed in a furnace and heated at 800°C for 2 days, after that it was slowly cooled down at rate 3°C per hour to 550°C, kept for 3 days, then slowly cooled down to room temperature. The product was cleaved easily along the basal plane. The powder of $Bi_2Se_3$ ground from single crystal was identified by X-ray powder diffraction.

We measured the resistance of $Bi_2Se_3$ single crystal at high pressure by four-probe methods in a diamond anvil cell (DAC) made of CuBe alloy as described in Ref. 18, 19. To avoid the electrode leads contacting with the metallic gasket, cubic boron nitride (cBN) fine powders were covered on the gasket made of T301 stainless steel. The electrodes are slim Au wires with 18μm in diameter. The gasket, preloaded from the thickness of 250μm to 60μm, was drilled a hole of 250 micrometer diameter wherein cBN insulating layer was pressed to place the transmission medium hexagonal BN (hBN). The culet of the diamond anvil is about 500 micrometer. The dimension of $Bi_2Se_3$ single crystals for resistance experiments was about 70μm*70μm*10μm with soft hBN fine powders around as pressure transmitting medium. The pressure was determined by ruby fluorescence method.[20,21] The diamond anvil cell was put inside a MagLab system to perform the experiments. To make sure the equilibrium of temperature, the temperature decreased slowly and was automatically controlled by the MagLab system. A thermometer nearby the diamond in the diamond anvil cell monitored the sample temperature accurately. The Hall coefficient at high pressure was measured by using Van der pauw method.

The X-ray diffractions with synchrotron radiation at high pressure and low temperature using a symmetric Mao Bell diamond anvil cell were done at the HPCAT of Advanced Photon Source (APS) of Argonne National Lab (ANL) with a wavelength 0.398Å. The two-dimensional image plate patterns are converted to one-dimensional 2Θ versus intensity data using Fit2D software package.

$Bi_2Se_3$, whose structure is the same as that of $Bi_2Te_3$, has a nearly idealized single Dirac cone and the largest band-gap (approximately 0.3eV) among aforementioned 3D TIs.[6, 7] $Bi_2Se_3$ is a direct gap semiconductor,[13] but strong spin-orbit coupling inverts the bulk bands at Γ point.[6] There are some previous reports about effects of pressure on $Bi_2Se_3$.[22, 23] Here we measured the transport property of $Bi_2Se_3$ single crystal in diamond anvil cell (DAC) via increasing the pressure for several times using different specimens. The behaviors of $Bi_2Se_3$ under high pressure are highly repetitive. The carrier density for the $Bi_2Se_3$ single crystals is around $2*10^{18} cm^{-3}$ at ambient pressure when the temperature is 30K. Since the vacancy of selenium causes Fermi surface above the bottom of conduction-band, $Bi_2Se_3$ is of n type carriers.

Figure 1 presents the temperature dependence of resistance at different pressures over an extensive range, measured in the a-b plane while monotonically increasing pressure. Figure 1(a) shows the temperature dependence of resistance at ambient pressure phase with small pressure interval. When pressure is below 3.1GPa, $Bi_2Se_3$ displays weak metallic behavior. It becomes more metallic above 5.1GPa, which is probably related to the pressure-induced electronic topological transition (ETT) in $Bi_2Se_3$ near 5GPa as reported.[23] The behavior change of the sample at ambient phase is similar to the report in the Ref. 22. To confirm the resistance behavior is repeatable, we measured specimens from different batches as shown figure (b) and (c), respectively. Figure (b) illustrates the resistance versus temperature with big pressure interval. The general trend of the resistance is metallic, semiconducting and metallic, which is the same to the results of figure 1 (a). More interesting, we observed clear superconducting transitions among several experiments. It is found that resistance drop firstly occur around 12GPa with the onset temperature (Tc) at about 4.4K. The Tc is defined using the same method described in Ref. 18, which is based on the differential of resistance over temperature. The selenium vacancy concentration can make the resistance and carrier density at ambient pressure vary by several orders of magnitude.[24] Different initial carrier density could be the origin of the somewhat different temperature dependences of the resistance from Ref22. We assume that a relative higher carrier density such in the level above $10^{18}cm^{-3}$ would be crucial to generate superconductivity in $Be_2Se_3$. The evolution of resistance as a function of temperature at low temperature from 12.5GPa to 31.9GPa is shown in figure 1(c). The onset temperature with a clear resistive drop fleetly increases to the highest temperature of 8.5K, then decreases, and then keeps almost a constant by further increasing the pressure.

To assure whether the resistive drop is indeed a superconducting transition, we measured the transition temperature at variant external magnetic fields. Figure 2 shows the measured resistance in low temperature range at 23GPa with applied magnetic field H perpendicular to the a-b plane of single crystal (a). Increasing magnetic field, the onset temperature shifts toward lower temperature and finally zero resistance is lost at higher magnetic field, which is strong evidence that the transition is superconductivity in nature. The change of magnetic field H with superconducting transition temperature ($T_c$) is shown in the figure 2 (b). Using the Werthdamer-Helfand-Hohenberg formula[24], $H_{c2} = -0.691 \times [\frac{dH_{c2}}{dT}]_{T=T_c} \times T_c$ , the upper critical field $H_{c2}(0)$ is extrapolated to be 4.7T for H//c axis at 23GPa.

To demonstrate the results are highly repetitive, we put results from different measurements on several specimens together. Figure 3 shows the $T_c$ and the carrier density dependence on pressure. The evolution of $T_c$ is shown in figure 3 (a). According to the results from figure 3 (a), we divide the range of the pressure into two regions (ambient phase and high pressure phases). Increasing the pressure from 12.5GPa to 31.9GPa, $T_c$ increases quickly to the highest temperature (around 8.5K), then decreases to near 6.5K when pressure increases to around 21GPa, and then keeps almost a constant. Electrical transport and optical measurement show $Bi_2Se_3$ is a n-type doped semiconductor.[24, 26-28] The p-type behavior of $Bi_2Se_3$ has been induced through slight chemical substitution.[29, 30] The electron state changes from the hole-dominated to electron-dominated type in $Bi_2Te_3$ induced by pressure and the $T_c$ dependence on pressure of $Bi_2Te_3$ is closely related to the change of the differential of hall coefficient over pressure.[31] In order to study the carrier density change of $Bi_2Se_3$ we performed the Hall effect measurements from 2GPa to 23GPa with a magnetic field perpendicular to the a-b plane of the sample up to 7T, by sweeping the magnetic field at a fixed temperature (30K) for each given pressure. The Hall coefficient measurements indicate the carrier of crystals used in the experiments is electron-dominated type which do not change over the entire measured pressure range. From several measurements at ambient pressure, we calculated the carrier density to be approximately $10^{18}/cm^3$. From 2GPa to 23GPa, the carrier density increases by more than two orders of magnitude at temperature 30K as shown in figure 3 (b). Besides, the carrier density increases abruptly around 5.5GPa (ETT pressure) and 12GPa (superconducting onset pressure). It changes quickly at high pressure phase. Meanwhile, when the carrier density goes up to the maximum value, $T_c$ is the highest too. Therefore, the variation of $T_c$ versus pressure is closely connected with the change of carrier density at pressure.

To investigate the crystal structure evolution as function of pressure, we conducted high pressure x ray diffractions with synchrotron radiation at 9K that is a temperature range near the superconducting transition. At 9 K and pressures below 9.6GPa, $Bi_2Se_3$ remains ambient phase, but a high pressure phase sets in with further increasing pressure. Phase transitions takes place at around 12GPa and 20GPa, respectively, as shown in figure 4 (a). Figure 4 (b) shows the lattice parameter a, c and volume V versus pressure at ambient pressure phase, which decrease with increasing pressure.

Referring to results of synchrotron radiation with Ref. 22, we infer that $Bi_2Se_3$ has four phases: ambient phase I from 0 to 12 GPa with rhombohedral R-3m structure; high pressure phase I (12~20GPa), high pressure phase II (20~30GPa), and high pressure phase III (>30GPa) with monoclinic sevenfold C2/m structure, monoclinic eightfold C2/c structure and BCC Im-3m structure, respectively.

Figure 5 shows the global phase diagram of n-type $Bi_2Se_3$ single crystal as function of pressure up to 32GPa. The black and blue balls represent independent experimental data. According to the crystal structure phase transition at 9K at pressures of around 12GPa, 20GPa and >29GPa, respectively, the phase diagram is composed of four areas. $Bi_2Se_3$ becomes a superconductor at its high pressure phases of HP I, HP II and HP III.

In summary, we found the superconductivity of $Bi_2Se_3$ at its high pressure phases. The $T_c$

dependence of pressure is closely dependent of change of carrier density. The x ray diffractions with synchrotron radiation further indicate four crystal structures of $Bi_2Se_3$ as function of pressure with high pressure states being superconductive.

**Acknowledgments**

This work was supported by NSF & MOST through research projects. HPSynC is supported as part of EFree, an Energy Frontier Research Center funded by the U.S. Department of Energy under Award DE-SC0001057. HPCAT is supported by DOE-BES, DOE-NNSA (CDAC), and National Science Foundation

**Figure captions**

**FIG.1.** (Color) The resistance of $Bi_2Se_3$ single crystal as function of temperature at ambient pressure phase (a) (with small pressure interval), (b) (with large pressure interval) and high pressure phase showing superconductivity (c). The first superconducting pressure is around 12GPa. (a) and (b) are independent experiments.

**FIG.2.** (Color) The superconducting transition of $Bi_2Se_3$ with applied magnetic field H perpendicular to the a-b plane of single crystal at 23GPa (a). The $T_c$ dependence of the magnetic field is showed in (b), and the upper critical fields $H_{c2}(0)$ are extrapolated to be 4.7T at 23GPa.

**FIG.3.** (Color) The Tc (a) and the carrier density (b) as function of pressure from independent experiments.

**FIG.4.** (Color) The x-ray diffraction spectra of $Bi_2Se_3$ at different pressure and T=9K (a). The wavelength of synchrotron radiation is 0.398Å. (b) shows the lattice parameter a, c and volume V versus pressure at 9K.

**FIG.5.** (Color) The phase diagram of n-type $Bi_2Se_3$ single crystal as function of pressure up to 32GPa. The black and blue balls represent superconducting transition temperature from different experiments.

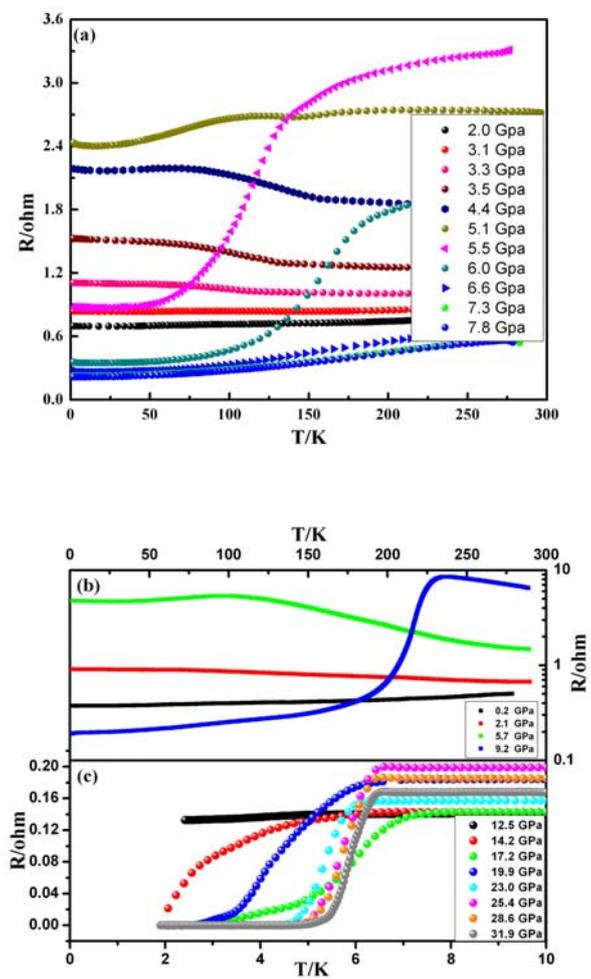

FIG.1.

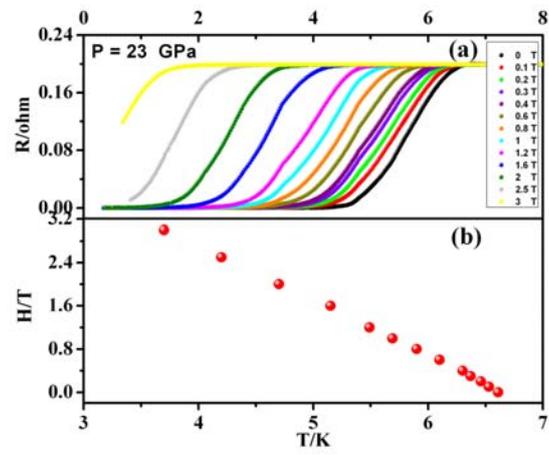

FIG.2

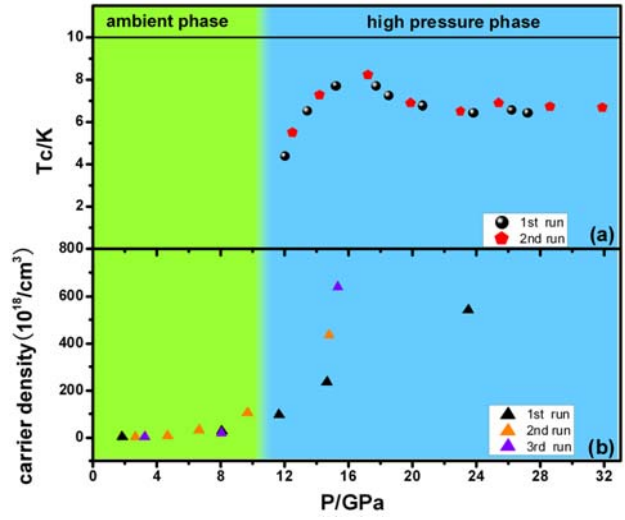

FIG.3

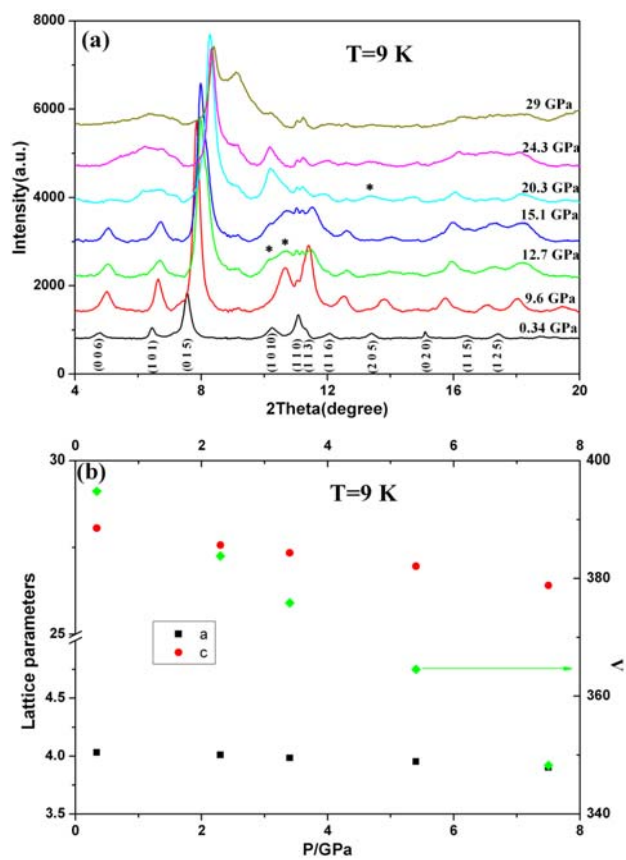

FIG.4

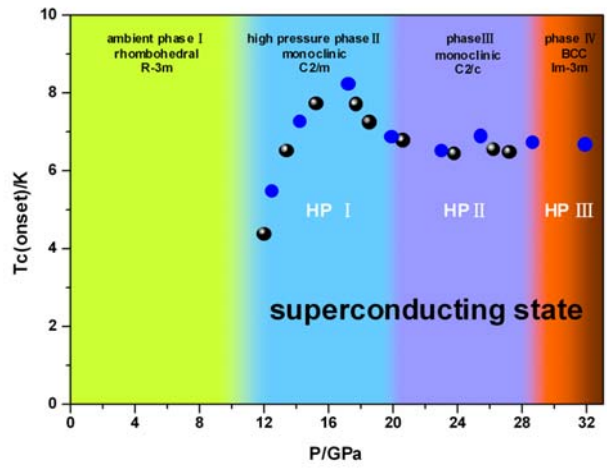

FIG.5.